# *Nemo*: a computational tool for analyzing nematode locomotion

George D. Tsibidis[1§] and Nektarios Tavernarakis[2]

[1]Institute of Electronic Structure and Laser, Foundation for Research and Technology, P.O. Box 1385, Vasilika Vouton, 71110 Heraklion, Crete, Greece
[2]Institute of Molecular Biology and Biotechnology, Foundation for Research and Technology, P.O. Box 1385, Vasilika Vouton, 71110 Heraklion, Crete, Greece

[§]Corresponding author

Email addresses:
    GDT:   tsibidis@iesl.forth.gr
    NT:     tavernarakis@imbb.forth.gr

# Abstract

The nematode *Caenorhabditis elegans* responds to an impressive range of chemical, mechanical and thermal stimuli and is extensively used to investigate the molecular mechanisms that mediate chemosensation, mechanotransduction and thermosensation. The main behavioral output of these responses is manifested as alterations in animal locomotion. Monitoring and examination of such alterations requires tools to capture and quantify features of nematode movement. In this paper, we introduce *Nemo* (<u>ne</u>matode <u>mo</u>vement), a computationally efficient and robust two-dimensional object tracking algorithm for automated detection and analysis of *C. elegans* locomotion. This algorithm enables precise measurement and feature extraction of nematode movement components. In addition, we develop a Graphical User Interface designed to facilitate processing and interpretation of movement data. While, in this study, we focus on the simple sinusoidal locomotion of *C. elegans*, our approach can be readily adapted to handle complicated locomotory behaviour patterns by including additional movement characteristics and parameters subject to quantification. Our software tool offers the capacity to extract, analyze and measure nematode locomotion features by processing simple video files. By allowing precise and quantitative assessment of behavioral traits, this tool will assist the genetic dissection and elucidation of the molecular mechanisms underlying specific behavioral responses.

# Background

*Caenorhabditis elegans* is a small (1 mm) free-living nematode worm that offers unique advantages for investigating fundamental problems in biology. The developmental and anatomical characterization of this animal is unparalleled in the metazoan world. The complete sequence of cell divisions that occur as the fertilized egg develops into the 959-celled adult are known [1, 2]. Furthermore, the description of neuronal connectivity in *C. elegans* is exceptionally detailed. Serial section electron microscopy has identified the pattern of synaptic connections made by each of the 302 neurons of the animal (including 5000 chemical synapses, 600 gap junctions, and 2000 neuromuscular junctions), so that the full "wiring diagram" of the animal is known [3, 4]. Indeed, the neural circuit of C. elegans shows the characteristics of a "small-world" network [5]. Despite such a simple nervous system, animals display a rich repertoire of behaviors including elaborate responses to chemical, mechanical and thermal cues and specific locomotory patterns [6, 7]. *C. elegans* is the first metazoan for which the genome was sequenced to completion [8]. The wealth of information on the biology of the organism, coupled with the broad range of genetic and molecular techniques applicable in *C. elegans*, allows in-depth studies of how genes specify and control neuron function to generate behavior [7, 9]. To this end, computational tools that facilitate the detailed analysis of nematode locomotion are highly desirable.

*C. elegans* sinusoidal locomotion ensues from alternate contraction and relaxation of dorsal and ventral body wall muscles, which generates a canonical sinusoidal pattern of movement [3, 9]. The arrangement of the body wall muscles and their synaptic inputs restricts locomotion to dorsal and ventral turns of the body. While, in principle, movement of the worm body is restricted in a two-dimensional space and it resembles a sinusoidal wave, many factors could affect its behavior. Numerous mutations disrupt normal sinusoidal locomotion in *C. elegans*, resulting in animals with movement defects ranging from total paralysis, to severe uncoordination, to subtle and almost imperceptible irregularities in movement [10-12]. As a result, the rate and direction of movement and the shape of its trajectory may change dramatically leading to more complex patterns. Some other factors affecting locomotion are the processes of feeding, egg-laying and mating, environmental stimuli, animal age and treatment with chemical substances [13-15].

While in some cases, behavioral alterations pertaining to animal movement are pronounced and may easily be described qualitatively, frequently such alterations are rather subtle or even imperceptible by simple observation. Thus, to obtain a better insight into a variety of behavioral effects and elucidate the mechanism governing the underlying processes, a more systematic analysis is required through approaches that provide precise quantitative information.

Various automated systems have been described aiming to track single or multiple worms and to study quantitatively their locomotion and behavior [13, 16-20, 23-26]. These systems offer the capacity to calculate global direct measurable parameters such as position of the animals and movement paths or indirect parameters such as speed, change in direction, shape, wavelength, and amplitude. In principle, analysis is restricted to the location of the head and the tail while the rest of the body is not investigated thoroughly [13, 16-20, 23-26]. Some of the systems are designed to observe and analyze locomotion of multiple animals at low magnification [13, 24-26]. However, because in these systems observation is conducted at low magnification, the detailed path of animal movement cannot be studied. Alternative computer tracking methods have been developed to overcome these constraints in which higher magnification is used. To maintain the animal within view, systems are equipped with a tracking program designed to control the movement of a motorized stage to keep the worm in sight [18]. The accuracy of the information obtained depends on the mechanics of the system and the integration with the microscope and camera. Other systems are used to examine more complicated behaviour, involving bends and reversals, however, only video sequences with worms in sight are analysed [20]. Most of the computer-driven systems [13,16-20,23-26] perform an automatic tracking and feature extraction without allowing the user to intervene and define regions of interest, set thresholds, accept or reject information, process data easily, or modify the computer algorithm and they usually assume a deep knowledge of programming languages. Additionally, the majority of the methods produce data that need to be interpreted independently and do not yield a complete picture of animal behaviour. To our knowledge, only one system has been developed [23] that provides a Graphical User Interface (GUI) to assist in a more comprehensive analysis. It is designed to control tracking and recording of the animal and subsequently illustrate the progress of the recognition process rather than offer the basis for a systematic quantitative analysis of the locomotionary behaviour.

In this paper, we present *Nemo*, a computational tool for obtaining quantitative information about nematode movement. This tool is designed to track deformable objects from a video sequence in high resolution. The algorithm we developed initially extracts morphological features and proceeds with segmentation of the animal body, retrieving information related to the position of the centre of mass of each body section separately. Segmentation allows movement details, body thickness and other information about any section of the worm to be easily acquired with minimum intervention from the user. A routine has also been integrated to compute the displacement of the image due to movement of the camera in order to keep the animal in sight. A particular advantage of our system is that it allows the user to choose regions of interest and compute specific locomotion features, related to these regions. This enables viewing image information for any part of the animal in the form of plots and histograms depicting the magnitude of particular movement parameters. Thus, features such as the wavelength, amplitude and direction change can be calculated in regions of interest.

In addition we have developed a GUI that automates the analysis and enables researchers to collect movement data accurately. While, the algorithm has originally been built for the study of simple behaviors such as the sinusoidal locomotion of the animal it can be readily generalized to process and describe more complex movement patterns.

**Data acquisition**

We followed standard procedures for *C. elegans* culture and maintenance [21]. The strain utilized in this study was the wild type Bristol isolate N2. Nematode rearing temperature was kept at $20^{o}$C. Animals were grown on agar plates with nematode growth medium (NGM), seeded with the bacterium *Escherichia coli* as a food source. For videotaping, 5-10 gravid adult worms were placed on fresh, seeded NGM plates and allowed to move freely for 30 min before observation. Animals were imaged via a Zeiss Stemi SV11 stereomicroscope (Carl Zeiss, Jena, Germany) with a Moticam 2000 CCD camera (Motic Instruments Inc., Richmond, Canada), at a resolution of 800x600 pixels and a frame rate of 40 fps. Video files of moving nematodes were generated using the software package accompanying the camera (Motic Images Plus 2.0; AVI, audio video interleave format).

# Implementation

**Algorithm**

Individual *C. elegans* animals in video frames are extracted by using routine image analysis techniques. Before image processing algorithms are applied, all indexed images are converted into grey images (Fig.1A). A mean low-pass filter is then applied to smooth each grey image. This filter replaces every pixel with the average of its 3x3 neighborhood. Every image is subsequently thresholded and converted into a binary image in which objects are separated from the background by clear boundaries. To avoid time consuming computations, after the initial extraction of the area occupied by the animal, processing is restricted only to small boxes containing the animal rather than to the entire image.

To extract useful quantitative information about individual animals in a frame, a number of morphological operations are applied on the binary images [22]. Dilation is one of the basic operators used and the basic effect of the operator is to gradually enlarge the boundaries of regions of foreground pixels. For labeling connected components in the image, every frame is scanned and pixels are grouped into components based on pixel connectivity. Large objects are assigned to animals in the image while smaller objects outside the perimeter of the animals are removed. Using built-in functions provided by the Image Analysis Toolbox of Matlab, the perimeter of each worm can be easily obtained in addition to the 'spine' (or 'skeleton') of the animal. We have also developed an algorithm to remove small and redundant 'branches' on the skeleton. The image processing procedure is summarized in Fig.2 A more detailed description of the algorithm can be found in supplementary data [see Additional file 1].

To analyze animal locomotion, the coordinates of points along the spine are computed by dividing the worm into a number of segments $N$ (in this experiment $N=7$ as seen in Fig.1A). The center of mass of every segment, as well as the centroid of the whole animal are recorded. The procedure followed to assign the anterior and posterior parts of the worm is based on the calculation of the smallest distances between the endpoints in two successive frames. The thickness of the animal is computed for all segments except from the head and tail.

To obtain more accurate quantitative information, locomotion is observed at high resolution. This requires a displacement of the camera to keep the animal in sight. A reference to an absolute coordinate system implies that reference points which operate as beacons must be introduced on the plate where the worm moves. Our proposed system is equipped with a routine that aims to correct the coordinates of every point of the animal due to camera displacement.

**Data Management**

A Graphical User Interface (GUI; Fig.1G) has been developed as an integral part of *Nemo*. The interface is designed to assist the analysis of collected data by operating as an information management tool, which allows the user to conduct further calculations. A detailed description of the GUI can be found in supplementary data [see Additional file 1].

The GUI provides a collection of tools managing specific actions. These tools facilitate operations such as:

- Highlighting morphological features (perimeter, area, centroids, trajectory), within images of the animal (Fig.1A).
- Generating graphs and histograms of measurable and computed parameters (distance covered by the worm, distance between the head and the tail, wavelength and amplitude of the sinusoidal movement, speed and direction change of the major axis of the animal) as shown in Fig.1B-F.
- Information management (analysis of only specific regions of interest, calculation of pertinent parameters, interactively).

In addition, the GUI provides the flexibility of easily adding new functionality (introduction of appropriate new tool buttons) beyond what is implemented in the current version.

## Results and discussion

We introduce a computationally efficient and robust system for analyzing nematode movement, which can be employed for quantitative, time-resolved studies of *C. elegans* behavior. The system is designed to be flexible in terms of functionality, ease of use and implementation. The overall goal of the software is to support fast, automated and semi-automated analysis of a large amount of video information, with high reliability and accuracy. The software also offers the capacity of focusing and analyzing specific image areas containing individual animals, which further enhances performance. The running time of the algorithm is determined primarily by the number of animals present in the frames and the complexity of the movement.

*Nemo* offers a rapid data extraction and overview and requires only minimum intervention by the user for particular calculations such as head-tail determination, thickness computation. The identification of the worm head and tail was performed by allowing the user to define initially the locations of the two ends and subsequently an automated algorithm was used to distinguish the head from the tail. A similar approach was pursued by another system [16-18] which assumes the head moves more frequently than the tail and that the tail area is darker than the head. User intervention provided by *Nemo* helps avoid confusion in the identification due to intensity differences or incorrect calculation of endpoint speed. The simple, sinusoidal type of locomotion ensures excellent accuracy of such an algorithm.

The system we present here measures the body thickness of every section of the animal based on an initial user intervention to define a circle around the centre of mass of every section of the body and the four intersection points with the perimeter can help to compute easily and faster the distance [see Additional file 1]. Other systems measure thickness at the centre, head and tail locations [16-18], first considering 9-pixel-long segments from the skeleton list, and then computing the best fit line for the segment. This approach requires relatively high resolution to provide an adequate number of points for analysis and is also time consuming.

Related systems focus on automated extraction of measurable quantities and morphological information [13, 16-20, 23-26]. The system we developed provides additional functionality by incorporating a GUI. This unique interface aims to offer a systematic quantitative analysis of *C. elegans* locomotary behaviour. Previous efforts [23] focus entirely on observing the animal skeleton at various time time points. By contrast, the GUI is incorporated into *Nemo* allows interactive user intervention and analysis of specific areas of image figures and data plots. Data viewing and graphing is handled by the GUI interactively.

We developed the system using Matlab, an easy to learn and implement algebraic package which is platform-independent [see *Availability and requirements* section). The versatility of the system allows easy expansion by incorporating new functions into the existing algorithm without need for major modifications to the code. Most of the previous algorithms were developed in C, Visual Basic or older computing languages [13, 16, 18-20, 24-26] which makes code more complicated and difficult to incorporate additions or updates. Two other automated systems [17, 23] written in Matlab, do not allow user intervention which limits the capacity to change or handle data. The system described here (main algorithm and GUI) facilitates the straightforward processing of video data and does not require extensive computer programming knowledge. The incorporated GUI allows the user to highlight morphological features, generate graphs and images and analyse information. Thus, the system is easy to use and readily accessible to the broad community of researchers.

In this study, *Nemo* is employed to process video sequences of single wild type worms crawling on plates. The algorithm can be readily generalized to handle a large number of animals and describe more complex behavioral and locomotion phenotypes by capitalizing on the convenience of the versatility of the system. This can be accomplished by simply introducing new arbitrary movement parameters for processing. To further improve the flexibility of the system we intend to develop a web based, version of the graphical user interface that will be posted online.

# Conclusions

We present, *Nemo*, an algorithm designed to measure and analyze nematode movement features by processing video image sequences. The system described here provides a powerful means of data extraction from 2D images. In conjunction with a GUI, *Nemo* constitutes an integrated approach to study nematode locomotion quantitatively by processing specific movement parameters and displaying measurable quantities. By enabling processing and reliable analysis of large amounts of data with high accuracy this system facilitates the systematic study and description of nematode behavior. While we only examined sinusoidal wild type animal movement to demonstrate the capacity of the tool, it can readily be utilized to handle complicated locomotion behaviors of both wild type and mutant animals, by introducing additional movement characteristics subject to quantification.

# Availability and requirements

Project name: Analyzing *C.elegans* locomotion
Project home page: http://elegans.imbb.forth.gr/
Operating system(s): *Nemo* was compiled and tested on a PC with 1.7GHz Pentium 4 running Windows XP. It was also run on Linux and Solaris platforms without any reported bug.
Programming language: Matlab (Release 13 or higher), AVI reader and player. The free program *mplayer* [http://www.mplayerhq.hu] was used to split the video sequence in individual frames, in JPEG format. Morphological data captured by *Nemo* are stored either as binary files or in text (TXT) format.
Licence: Matlab (The Math Works, Inc, Natick, US)
Any restrictions to use by non-academics: *Nemo* is a software tool free for academic and non-profit use.

The steps which are followed to extract all information required for quantitative analysis are described in supplementary data [see Additional file 1; http://elegans.imbb.forth.gr/Tsibidis-Nemo/Steps_ensued_for_quantitative_analysis]. The source code is freely available [see Additional file 2; http://elegans.imbb.forth.gr/Tsibidis-Nemo/Computer_algorithms]. Updates and bug fixes will be posted regularly on [http://elegans.imbb.forth.gr/Tsibidis-Nemo]

# Authors' contributions

GT and NT designed research, GT wrote the software, GT and NT analyzed data and evaluated results. GT and NT wrote the manuscript. All authors read and approved the submitted manuscript.

# Acknowledgements

We thank Giannis Voglis and Nikos Kourtis for help with data acquisition. The nematode strain used in this work was provided by the *Caenorhabditis* Genetics Center, which is funded by the NIH National Center for Research Resources (NCRR). This work was funded by grants from EMBO and the EU 6th Framework Programme to N.T and G.D.T.

# Figures

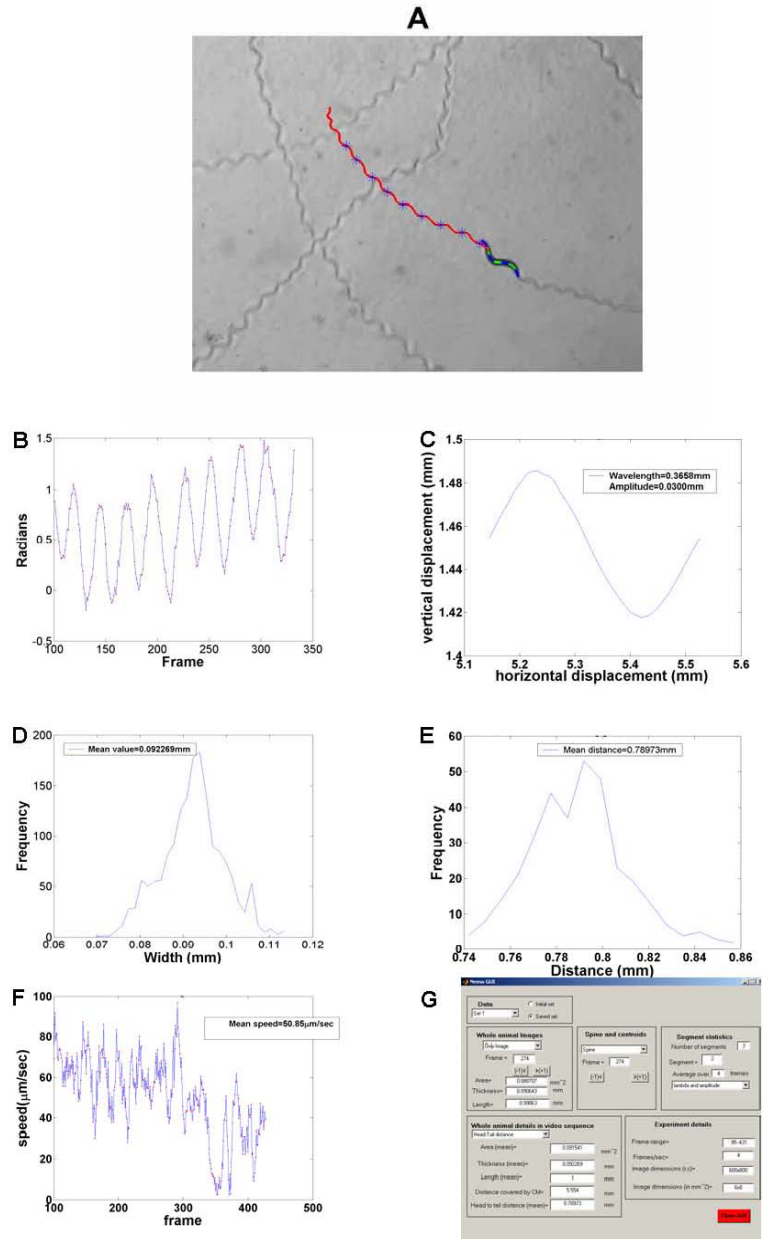

**Figure 1 - Images and data plots**

(A) Trajectory of the animal locomotion, (B) angle evolution with respect to frame number, (C) waveform of the movement, (D) Width histogram, (E) Distance between head-tail, (F) Speed series with respect to frame number, (G) Graphical User Interface

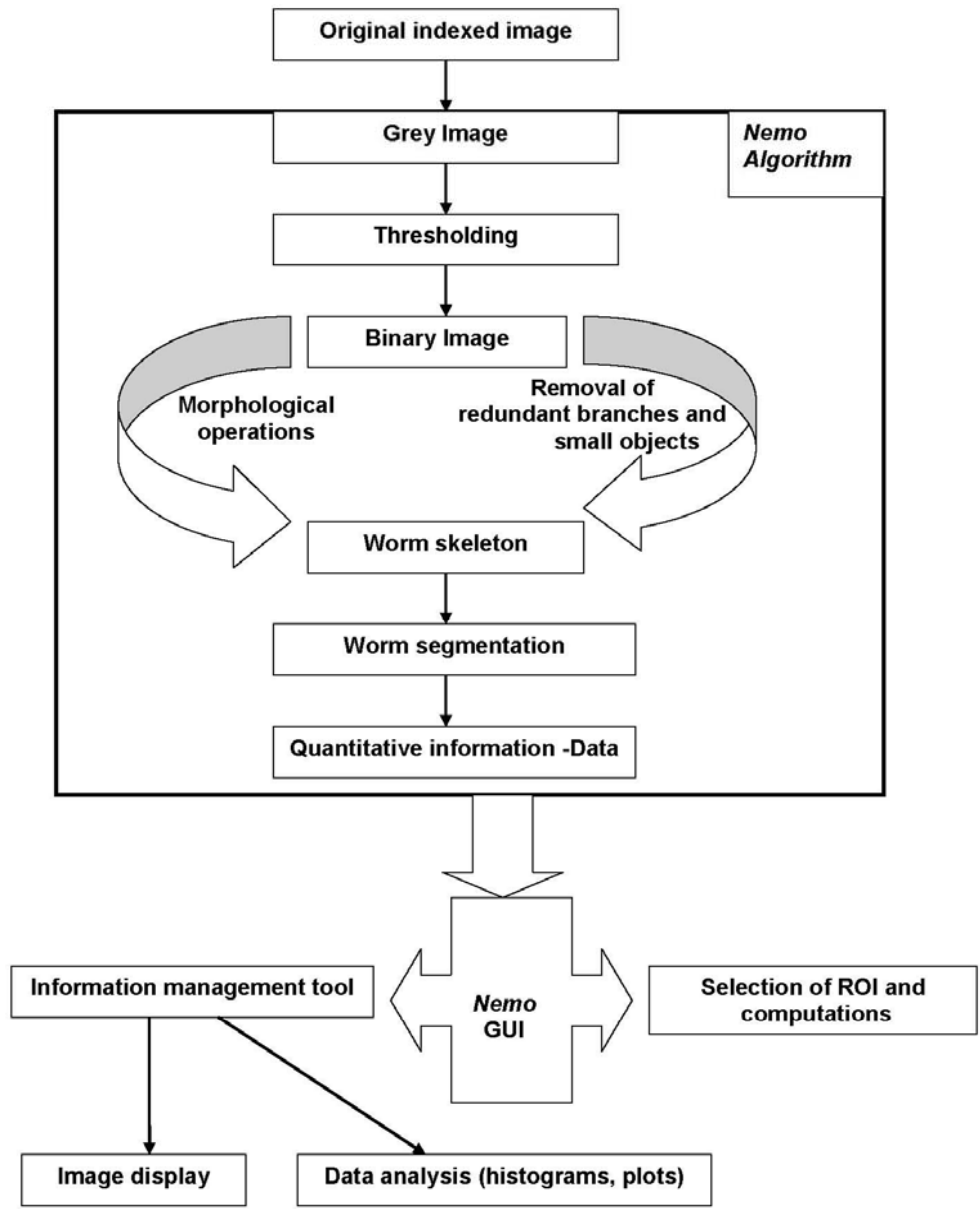

**Figure 2 - The *Nemo* algorithm**